\documentclass[physrev,reprint,superscriptaddress,bibnotes]{revtex4-2}  

\usepackage[T1]{fontenc}
\usepackage[australian]{babel}
\usepackage[a4paper,centering,hmargin=1.75cm,vmargin=2cm]{geometry} 
\usepackage{amsmath,amssymb,graphicx,bm,microtype}
\usepackage[dvipsnames]{xcolor}
\usepackage{booktabs}
\usepackage[colorlinks,allcolors=blue!50!black]{hyperref} 
\usepackage[all]{hypcap}
\usepackage{cleveref}
\usepackage{braket}
\usepackage[italic,thinc]{esdiff} 
\usepackage[version=4]{mhchem} 
\usepackage{siunitx} 
\renewcommand{\tabcolsep}{4pt}

\newcommand{\dg}{\dagger}

\begin{document}

\title{Simulating open-system molecular dynamics on analog quantum computers}

\author{Vanessa C.~Olaya-Agudelo}
\thanks{These authors contributed equally to this work.}
\affiliation{School of Chemistry, University of Sydney, NSW 2006, Australia}

\author{Ben~Stewart}
\thanks{These authors contributed equally to this work.}
\affiliation{School of Chemistry, University of Sydney, NSW 2006, Australia}

\author{Christophe H.~Valahu}
\affiliation{School of Physics, University of Sydney, NSW 2006, Australia}
\affiliation{ARC Centre of Excellence for Engineered Quantum Systems, University of Sydney, NSW 2006, Australia}

\author{Ryan J.~MacDonell}
\affiliation{Department of Chemistry, Dalhousie University, Halifax, NS B3H 4R2, Canada}

\author{Maverick~J.~Millican}
\affiliation{School of Physics, University of Sydney, NSW 2006, Australia}
\affiliation{ARC Centre of Excellence for Engineered Quantum Systems, University of Sydney, NSW 2006, Australia}

\author{Vassili G.~Matsos}
\affiliation{School of Physics, University of Sydney, NSW 2006, Australia}
\affiliation{ARC Centre of Excellence for Engineered Quantum Systems, University of Sydney, NSW 2006, Australia}

\author{Frank~Scuccimarra}
\affiliation{School of Physics, University of Sydney, NSW 2006, Australia}
\affiliation{ARC Centre of Excellence for Engineered Quantum Systems, University of Sydney, NSW 2006, Australia}

\author{Ting Rei~Tan}
\affiliation{School of Physics, University of Sydney, NSW 2006, Australia}
\affiliation{ARC Centre of Excellence for Engineered Quantum Systems, University of Sydney, NSW 2006, Australia}

\author{Ivan~Kassal}
\email{ivan.kassal@sydney.edu.au}
\affiliation{School of Chemistry, University of Sydney, NSW 2006, Australia}
\affiliation{ARC Centre of Excellence for Engineered Quantum Systems, University of Sydney, NSW 2006, Australia}

\begin{abstract}
Interactions of molecules with their environment influence the course and outcome of almost all chemical reactions. However, classical computers struggle to accurately simulate complicated molecule-environment interactions because of the steep growth of computational resources with both molecule size and environment complexity. Therefore, many quantum-chemical simulations are restricted to isolated molecules, whose dynamics can dramatically differ from what happens in an environment. Here, we show that analog quantum simulators can simulate open molecular systems by using the native dissipation of the simulator and injecting additional controllable dissipation. By exploiting the native dissipation to simulate the molecular dissipation---rather than seeing it as a limitation---our approach enables longer simulations of open systems than are possible for closed systems. In particular, we show that trapped-ion simulators using a mixed qudit-boson (MQB) encoding could simulate molecules in a wide range of condensed phases by implementing widely used dissipative processes within the Lindblad formalism, including pure dephasing and both electronic and vibrational relaxation. The MQB open-system simulations require significantly fewer additional quantum resources compared to both classical and digital quantum approaches.
\end{abstract}

\maketitle

Most molecular dynamics is governed by interactions between molecules and their surroundings. With the exception of molecules in vacuum or dilute gas, chemistry occurs in environments---from solvents to crystals to proteins---which decohere the molecule's quantum state and can transfer energy to and from the molecule. Molecular dynamics in condensed phase can differ completely from that in the isolated molecule because molecule-solvent interactions can induce processes such as non-adiabatic transitions, charge transfer, and barrier crossing~\cite{Nitzan}. 

Accurate simulations of an isolated molecule's quantum dynamics are challenging on conventional computers due to the exponential growth of the Hilbert space with molecular size. Extending such simulations to an open-system treatment is even more difficult and usually prohibitive, as it often involves representing the molecule's mixed state with a density matrix. These limitations have constrained the most accurate open-system simulation methods to small molecules~\cite{raab1999a,Domcke2002,RatnerLindblad2004,RatnerLindblad2005,HEOMPyrazine2016,HEOM2016method2}.


Quantum computers promise to simulate molecular quantum dynamics efficiently (in polynomial time in system size), but, like classical computers, usually require more resources to simulate open systems. Quantum computers can outperform classical ones by encoding the molecular quantum states on an inherently quantum platform~\cite{Kassal2008,Miessen2021,Lee2021,Miessen2023}. Nevertheless, current quantum algorithms for simulating open-system dynamics require more quantum resources (such as qubits and operations), making them prohibitively expensive for current quantum hardware. Two broad approaches have been pursued for representing system-bath interactions on quantum computers. In the first, additional qubits are added to represent the environment~\cite{digitalOQS2011Wang,OQSalgo2015,OQSalgo2016,digitalOQS2021}. In the other, intrinsic noise of the quantum device is used to mimic the environment~\cite{Tseng2000, Jose2023}. 
However, when the intrinsic noise does not directly correspond to the desired dissipative process~\cite{Tseng2000}---as is the case when qubit-based computers are used to simulate bosonic modes---using the intrinsic noise may be either impossible or require additional resources (whether qubits or gates) to transform the intrinsic noise into the desired form~\cite{Jose2023}. Hybrid approaches are also possible, which combine intrinsic noise on ancilla registers representing an environment~\cite{Juha2023,stadler2025}.
Therefore, in both approaches, and especially when simulating systems that cannot be directly mapped to qubits, considerable additional quantum resources are generally required to transform a closed-system quantum simulation into an open one. 

Analog quantum simulators using bosonic degrees of freedom promise to simulate molecular quantum dynamics with even fewer resources than digital quantum computers. An analog simulator is a purpose-built device whose Hamiltonian can be engineered to match that of the simulated system, so that the simulator's time evolution reproduces the desired dynamics.
For molecular simulations, analog simulators based on the mixed-qudit-boson (MQB) encoding represent molecular electronic and vibrational degrees of freedom using the qudit and bosonic modes of a trapped ion or a circuit quantum electrodynamics (cQED) system~\cite{MacDonell2021}. Using bosonic degrees of freedom to simulate nuclear motions---instead of encoding them in many qubits---reduces the quantum resource requirements by about an order of magnitude compared to qubit-based digital quantum simulation~\cite{MacDonell2021}. Experimental demonstrations of MQB simulation include simulations of conical intersections~\cite{Valahu2023,Whitlow2023,Wang2023} and vibronic spectra~\cite{MacDonell2023}. 


However, MQB simulations of molecular dynamics have not been fully generalised to open-system dynamics, with existing proposals limited to vibrational dissipation in trapped-ion devices~\cite{MacDonell2021}. Other theoretical~\cite{Mostame_2012,Lemmer_2018,Kim2022} and experimental~\cite{Gorman2018,Wang2024,so2024,sun2024} works on the quantum simulation of energy or charge transfer problems used bosonic degrees of freedom in open-system analog simulations, but were restricted to only some noise mechanisms or were not applicable to molecular quantum dynamics because bosonic modes were used only to represent the environment, not the system of interest.


Here, we present a framework for simulating open-system molecular dynamics with an MQB simulator. Our approach has two significant strengths. First, the dissipation can be engineered using both native dissipation and injected controllable dissipation, on both the qudit and the bosonic modes, so that chemical dissipative processes can be easily mapped onto the MQB simulator. Doing so turns decoherence mechanisms that usually limit analog simulations into a resource that makes the simulations more powerful. Second, engineering the dissipative processes requires few additional resources, whose number is often independent of the molecular size. In particular, we show how chemically relevant dissipative processes---including both electronic and vibrational relaxation and dephasing---can be implemented with existing trapped-ion technology. Overall, MQB simulators can solve the harder problem of open-system simulation with minimal overhead and with greater resistance to errors.

\section{Molecular open-system dynamics}
\label{sec:Lindbladsection}

There are many approaches to modelling open molecular systems, depending on where the molecule-environment boundary is drawn and what techniques are used to simulate either the molecule, the environment, or their interaction~\cite{Nitzan,MayandKuhn}. In all cases, open-system simulations aim to simulate the environment using a simpler method than that used for the molecule itself. The approaches to simulating the environment range from ones that explicitly describe the components of the environment, often at considerable computational cost, to those that treat either the environment or its influence on the molecule in an effective way.

A logical starting point for developing analog open-system approaches is the Lindblad master equation~\cite{Lindblad1976,Petruccione_2002}, the most general completely positive and trace-preserving Markovian master equation. It describes the dynamics of the molecule's reduced density matrix ${\rho}$ (we set $\hbar = 1$ throughout) by
\begin{equation}
    \diff{{\rho}}{t} = -i [ {H}^\mathrm{mol},{\rho} ] + \sum_i  \gamma_i^{\mathrm{mol}} \mathcal{D}\left[{L}_i\right]\rho,  
    \label{eq:LindbladME}
\end{equation}
where ${H}^\mathrm{mol}$ is the Hamiltonian of the isolated molecule, which can be specified in any form, including as a vibronic coupling (VC) model~\cite{Domcke,MacDonell2021} (for example, see the VC Hamiltonian in \cref{eq:closed-system-molecularH}). Each dissipative superoperator $\mathcal{D}\left[{L}_i\right]$ with rate $\gamma_i^{\mathrm{mol}}$ acts on $\rho$ as
\begin{equation}
    \mathcal{D}\left[L_i\right]\rho = \left({L}_i {\rho} {L}_i^{\dag} - \tfrac{1}{2}\{{L}_i^{\dag}{L}_i,{\rho}\}\right),
    \label{eq:LindbladRates}
\end{equation}
describing a dissipative process by a Lindblad operator ${L}_i$.

\begin{figure*}[t]
	\centering
	\includegraphics[width=\textwidth]{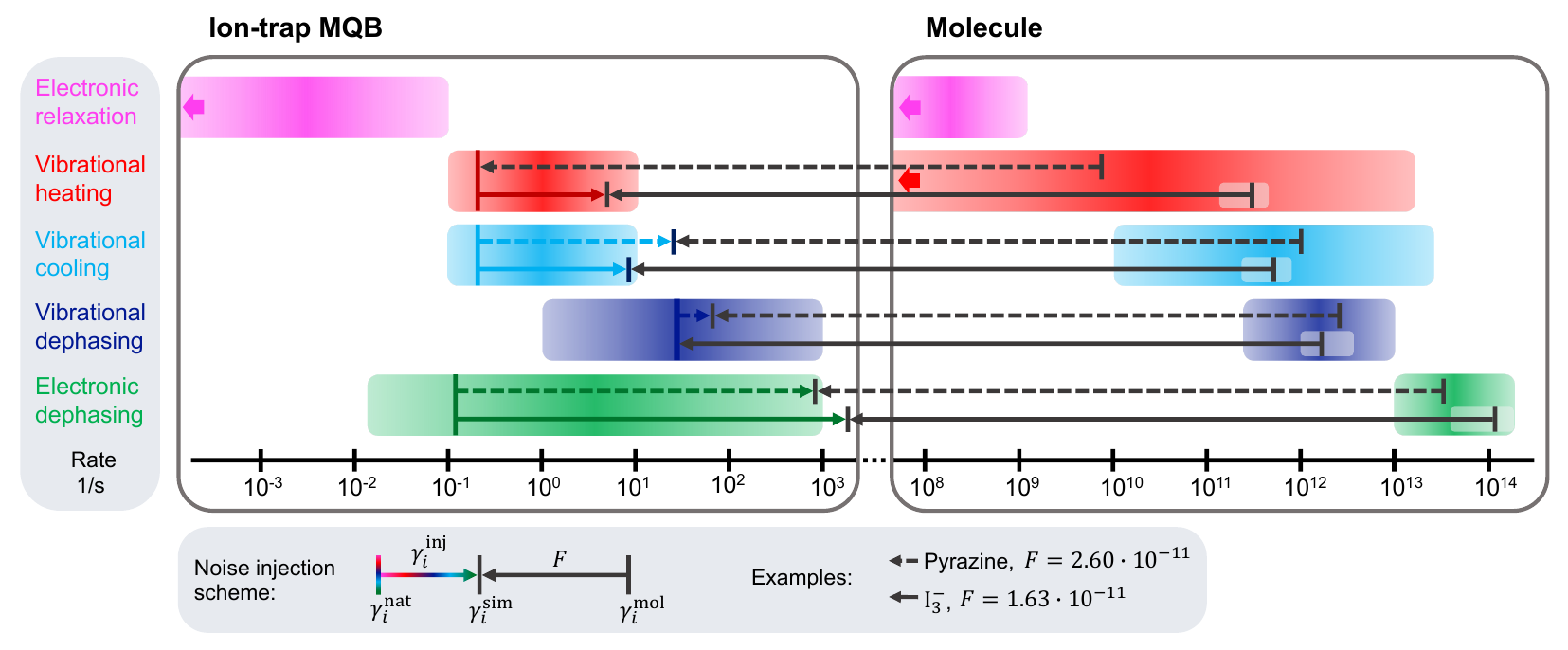}
	\caption{Typical rates for the most-relevant dissipative processes, \textbf{(left)}~in a trapped-ion MQB simulator~\cite{Ruster2016,MacDonell2023,lucas2007,Talukdar2016,Allcock2021,Jarlaud2021} and \textbf{(right)}~in molecules in condensed phase (modified from~\cite{Nitzan}). For both pyrazine~\cite{Schneider_1990,Schneider_1989,Stock_1990} and triiodide~\cite{Banin_1994} examples, each dissipative rate $\gamma_i^{\mathrm{mol}}$ (and range: grey highlights) is mapped onto the corresponding trapped-ion rate $\gamma_i^{\mathrm{sim}} = F \gamma_i^{\mathrm{mol}}$ using the scaling factor $F = \max_i \gamma_{i}^{\mathrm{nat}}/\gamma_{i}^{\mathrm{mol}}$ (black arrows). The maximum ratio is obtained for vibrational pure dephasing for triiodide and vibrational heating for pyrazine; therefore, no injection is required for these types of dissipation. Other dissipative processes require injected rates $\gamma_i^{\mathrm{inj}}$ (coloured arrows), to ensure that $\gamma_i^{\mathrm{sim}}=\gamma_i^{\mathrm{nat}}+\gamma_i^{\mathrm{inj}}$.}
	\label{fig:fig1}
\end{figure*}

In chemical contexts, there are usually four dominant dissipation mechanisms, which can be derived from microscopic principles. \Cref{eq:LindbladME} is usually derived from the related Redfield master equation, which is obtained from a perturbative treatment of the system-environment coupling in the Markovian limit~\cite{MayandKuhn,Nitzan}.
A Redfield equation is usually converted to a Lindblad equation by making the secular approximation~\cite{MayandKuhn,Petruccione_2002,Nitzan}, which is valid when the timescales of interest are slower than any rapidly oscillating terms in the master equation, allowing them to average out. In the secular limit, population relaxation and pure dephasing become decoupled~\cite{MayandKuhn,Nitzan}, making it possible to classify the most important dissipative processes in open chemical systems. With typical timescales given in \cref{fig:fig1}, they are:
\begin{enumerate}
    \item \textit{Radiative electronic relaxation} is the decay of excited electronic populations by spontaneous emission, including both fluorescence and phosphorescence. It is described by the dissipator $\mathcal{D}[\ket{n}\bra{m}]$ and rate $\gamma_{nm}^{\mathrm{mol}}$ for electronic states with energies $\varepsilon_m > \varepsilon_n$. Non-radiative electronic relaxation, such as intersystem crossing and internal conversion, are indirect consequences of vibrational relaxation and vibronic coupling, meaning that they are accounted for by the processes below. 
    
    \item \textit{Vibrational heating and cooling} are the gain and loss of energy in molecular vibrational modes, and are the most important way for a molecule to reach equilibrium with a thermal environment. For mode $j$, heating and cooling have dissipators $\mathcal{D}[a^\dg_j]$ and $\mathcal{D}[a_j]$, with rates $\gamma_{+,j}^{\mathrm{mol}}$ and $\gamma_{-,j}^{\mathrm{mol}}$, respectively. Detailed balance at temperature $T^{\mathrm{mol}}$ requires $\gamma_{+,j}^{\mathrm{mol}} = \gamma_{-,j}^{\mathrm{mol}} \exp(- \nu_j^{\mathrm{mol}}/k_\mathrm{B}T^{\mathrm{mol}})$ for mode frequency $\nu_j^{\mathrm{mol}}$.

    \item \textit{Electronic pure dephasing} is the reduction in phase coherence between electronic eigenstates, described by the dissipator $\mathcal{D}[\ket{n}\bra{n}]$ and rate $\gamma_{\mathrm{e},n}^{\mathrm{mol}}$ for electronic state $n$.
    
    \item \textit{Vibrational pure dephasing} is the decay of phase coherence between vibrational states, described by the dissipator $\mathcal{D}[{a}_j^{\dg}{a}_j]$ and rate $\gamma_{\mathrm{v},j}^{\mathrm{mol}}$ for vibrational mode $j$. 
\end{enumerate}

\section{Analog Open-System Simulation}
\label{sec:SimulationFramework}

An analog simulator evolves in time in the same way as the molecule being simulated, so that the molecule's dynamics can be retrieved by measuring the simulator evolution. To achieve this, the molecular Hamiltonian $H^{\mathrm{mol}}$ must be mapped onto the controllable simulator Hamiltonian $H^{\mathrm{sim}}$. To allow for a change of time scales, the mapping may include a scaling factor $F$ such that $H^{\mathrm{sim}} = F H^{\mathrm{mol}}$, where $F$ represents the ratio of the typical energy scale of the simulator to that of the molecule. Here, we extend the existing closed-system simulation approach to open-system ones and, by considering how this change affects the possible values of $F$, show that there is always a performance advantage in open-system simulations compared to closed ones.

For open-system simulations, two sources of dissipation can be used by the simulator: \textit{native} uncontrollable dissipation and \textit{injected} controllable dissipation. Native dissipation is intrinsic, always present in the simulator, and cannot be tuned. Conversely, injected dissipation consists of processes that can be engineered. Native dissipation is often considered to be a hindrance in quantum systems, but, in analog simulations, it can be a valuable tool for open-system simulations that extends the possible duration of simulations.


Native dissipation can be useful in an analog simulation if it contains a significant component that is Markovian and stable. Fortunately, as in molecules, dominant dissipative processes in analog hardware are usually well described using Lindblad dissipators, $\mathcal{D}[L_i^{\mathrm{nat}}]$, with corresponding rates, $\gamma_i^{\mathrm{nat}}$~\cite{whineland1998}. Indeed, if the coupling to the environment is weak and the secular approximation holds, the same four dissipative processes that are dominant in molecules will be dominant in the simulator. We assume that the simulator's dissipation rates are stable, i.e., that they can be measured in a preliminary experiment and remain constant throughout subsequent simulations. Stability is manageable because it only requires the dissipation to be stable over several runs of an experiment before it can be calibrated again.

We then refer to native dissipators $\mathcal{D}[L_i^{\mathrm{nat}}]$ as \textit{usable} if they implement a desired molecular dissipator $\mathcal{D}[L_i]$. In contrast, \textit{unusable} dissipation includes all native dissipators that have no analog in the molecule, as well as processes that cannot be described by the Lindblad formalism (e.g., non-Markovian ones).

A molecular dissipator, $\mathcal{D}[L_i]$, can be simulated on the analog simulator by ensuring that its dissipation rate is related to the corresponding molecular rate by the same scaling factor $F$ as above, $\gamma_{i}^{\mathrm{sim}} = F \gamma_{i}^{\mathrm{mol}}$. This condition can be met by using native dissipation with rate $\gamma_i^{\mathrm{nat}}$ and injecting additional dissipation of the same type at rate $\gamma_i^{\mathrm{inj}}$ that satisfies 
\begin{equation}\label{eq:inj}
    \gamma_i^{\mathrm{sim}} = \gamma_i^{\mathrm{nat}} + \gamma_i^{\mathrm{inj}} = F \gamma_{i}^{\mathrm{mol}}.
\end{equation}

For a closed-system simulation, $F$ can be chosen within hardware constraints, $F \in [ F^{\mathrm{cs}}_\mathrm{min}, F^{\mathrm{cs}}_\mathrm{max} ]$ (``cs'' for closed system). The lower bound, $F^{\mathrm{cs}}_\mathrm{min} = t_{\mathrm{mol}}/\tau_d^{\mathrm{cs}}$, is set by the ratio of the desired total simulation time in the molecule, $t_{\mathrm{mol}}$, to the coherence time of the simulator, $\tau_d^{\mathrm{cs}}$, after which the simulator is deemed insufficiently reliable. The upper bound, $F^{\mathrm{cs}}_{\mathrm{max}}$, is the ratio of the maximum simulator interaction that can be engineered in $H^{\mathrm{sim}}$ and the largest interaction of that type in $H^{\mathrm{mol}}$. Consequently, the longest molecular time that can be simulated on the specific hardware is $t_\mathrm{max}^{\mathrm{cs}} = \tau_d^{\mathrm{cs}} F^{\mathrm{cs}}_\mathrm{max}$.

Including dissipation gives a new range of $F$ for open-system simulation, $F \in [ F^{\mathrm{os}}_\mathrm{min}, F^{\mathrm{os}}_\mathrm{max}]$ (``os'' for open system). $F^{\mathrm{os}}_{\mathrm{max}}$ remains set by the hardware capabilities and is therefore unchanged, $F^{\mathrm{os}}_{\mathrm{max}}$ = $F^{\mathrm{cs}}_{\mathrm{max}}$. By contrast, $F^{\mathrm{os}}_{\mathrm{min}}$ must now meet two conditions. First, as in the closed simulation, $F$ is constrained by the open-system coherence time $\tau^{\mathrm{os}}_d$ of the simulator, $F \ge t_\mathrm{mol}/\tau^{\mathrm{os}}_d$. Because some dissipation is used in the simulation and $\tau^{\mathrm{os}}_d$ is determined only by the unused dissipation, $\tau^{\mathrm{os}}_d > \tau_d^{\mathrm{cs}}$, giving an improvement over the closed-system case. Second, because all $\gamma_i^{\mathrm{inj}}$ must be positive, \cref{eq:inj} requires that $F \gamma_i^{\mathrm{mol}} \geq \gamma_i^{\mathrm{nat}}$ for all $i$, that is, $F \ge R = \max_i\gamma_i^{\mathrm{nat}}/\gamma_{i}^{\mathrm{mol}}$. Overall, the lower bound for the open-system simulation is $F^{\mathrm{os}}_\mathrm{min} = \mathrm{max}(t_\mathrm{mol}/\tau^{\mathrm{os}}_d, R)$.

Importantly, using some of the native noise in the simulation always extends the maximum simulation duration. The longest possible open-system simulated time is $t^{\mathrm{os}}_\mathrm{max} = \tau^{\mathrm{os}}_d F^{\mathrm{os}}_\mathrm{max} = \tau^{\mathrm{os}}_d F^{\mathrm{cs}}_\mathrm{max}$; because $\tau^{\mathrm{os}}_d > \tau_d^{\mathrm{cs}}$, we find that $t^{\mathrm{os}}_\mathrm{max} > t_\mathrm{max}^{\mathrm{cs}}$. Therefore, any hardware that can be used for closed-system simulation can achieve longer simulated time when used for open-system simulation.

If $F^{\mathrm{os}}_\mathrm{min} = R$, then the $F$ that maximises the use of native dissipation and minimises injected dissipation is $F = F^{\mathrm{os}}_\mathrm{min}$; doing so means that no injected dissipation is required for the process with the maximum ratio $\gamma_i^{\mathrm{nat}}/\gamma_{i}^{\mathrm{mol}}$ (see \cref{sec:Examples} for two examples).

\section{Closed-System MQB Simulation}
\label{sec:MQBSimulator}

MQB simulators~\cite{MacDonell2021} are examples of analog quantum simulators, in which the molecular electronic and vibrational degrees of freedom are encoded in a qudit and multiple bosonic modes of the simulator. In a trapped-ion MQB implementation, the qudit is encoded in the electronic states of one of the ions, while the bosonic modes correspond to the collective normal modes of motion (vibrational modes) of the ion chain. MQB simulators are programmable because all parameters---including the energies of all states and the couplings between degrees of freedom---can be controllably adjusted using light-matter interactions~\cite{MacDonell2021}. 

The quantum resources required for an MQB simulation scale linearly with molecule size~\cite{MacDonell2021}. A chain of $N$ trapped ions contains $3N$ vibrational modes, meaning that a molecule with $d$ electronic states and $N$ vibrational modes can be mapped onto a single qudit in a chain of $\lceil{N/3}\rceil$ ions. 

The MQB approach implements vibronic-coupling Hamiltonians that are expressed as power series in the vibronic couplings~\cite{MacDonell2021}. The linear vibronic coupling (LVC) Hamiltonian, where the electronic and vibrational degrees of freedom are linearly coupled~\cite{MacDonell2021}, is
\begin{multline}
    H^{\mathrm{mol}} = \sum_{j}  \nu^{\mathrm{mol}}_j a^\dg_j a_j + \sum_{n,m} c_0^{(n,m)} \ket{n}\bra{m} + \\
    \sum_{n} \sum_{j \in \mathbf{t}} \frac{c_j^{(n)}}{\sqrt{2}} (a^\dg_j + a_j) \ket{n}\bra{n} + \\
    \sum_{n \neq m} \sum_{j \in \mathbf{c}} \frac{c_j^{(n,m)}}{\sqrt{2}} ( a^\dg_j + a_j ) \ket{n}\bra{m},
    \label{eq:closed-system-molecularH}
\end{multline}
where $\ket{n}$ are the electronic states, $a_j$ are the annihilation operators of the molecular vibrations, such that $Q_j= (a^\dg_j + a_j)/\sqrt{2}$ is the dimensionless position of the $j$th mode. The constants $c_0^{(n,n)}$ are the electronic energies, $c_0^{(n,m)}$ ($n\ne m$) are the inter-state couplings, $c_j^{(n)}$ and $c_j^{(n,m)}$ are the vibronic (vibrational-electronic) couplings for, respectively, the \textit{tuning} ($j \in \mathbf{t}$) and the \textit{coupling} ($j \in \mathbf{c}$) modes. Parametrising \cref{eq:closed-system-molecularH} requires carrying out an electronic-structure calculation in advance~\cite{MacDonell2021}. Higher-order vibronic-coupling terms could be readily included in $H^{\mathrm{mol}}$.

The molecular Hamiltonian, $H^\mathrm{mol}$, can be directly encoded on a trapped-ion simulator. The mapping begins with the Hamiltonian of a chain of $N$ ions~\cite{whineland1998},
\begin{equation}
    H^\mathrm{ion} = \sum_{j=1}^{3N} \nu_j a^\dg_j a_j 
    + \frac{1}{2}\sum_{n=0}^{d-1} \omega_n \ket{n}\bra{n},
    \label{eq:nolaserfield}
\end{equation}
where the vibrational mode $j$ has frequency $\nu_{j}$ and $\omega_n$ is the frequency of the $n$th electronic state relative to $\ket{0}$, the lowest state used in the simulation.

The remaining terms in $H^\mathrm{mol}$ are simulated by adding light-matter interactions that induce couplings in the simulator~\cite{MacDonell2021}. These interactions can be implemented through stimulated Raman transitions using pairs of non-copropagating laser beams that are both approximately detuned by $\Delta$ from an electronic excited state $\ket{e}$ outside of the qudit~\cite{whineland1998,Leibfried2003,Lee_2005}. The frequency difference between the two Raman beams, $\Delta\omega_\mathrm{L}$, can be tuned to implement all the necessary interactions: electronic coupling, vibronic tuning, and vibronic coupling. 

Electronic coupling between states $\ket{n}$ and $\ket{m}$ is implemented by setting $\Delta\omega_\mathrm{L} = (\omega_m - \omega_n) + (\chi_m - \chi_n)$, where $\chi_m- \chi_n$ is a frequency shift relative to the electronic transition. In the interaction picture with respect to \cref{eq:nolaserfield} and after a rotating wave approximation, the electronic coupling Hamiltonian between states $\ket{n}$ and $\ket{m}$ is~\cite{whineland1998,Lee_2005} 
\begin{equation}
    H^\mathrm{ion}_{\mathrm{e},{n,m}} = \Omega_{n, m} \ket{n}\bra{m} e^{- i (\chi_{m} - \chi_{n})t}+ \mathrm{h.c.},
    \label{eq:H_coupling_electronic}
\end{equation}
with interaction strength $\Omega_{n, m} = g_{n,e,A}^* g_{m,e,B}/2\Delta$, which can be adjusted by varying either the detuning $\Delta$ or the intensity-dependent light-matter couplings $g_{n,e,A}^*$ and $g_{m,e,B}$ of the beams coupling $\ket{n}$ to $\ket{e}$ and $\ket{m}$ to $\ket{e}$, respectively.

Vibronic tuning for state $\ket{n}$ is implemented by setting $\Delta\omega_\mathrm{L}$ close to the frequency of a vibrational mode $j$ ($\Delta\omega_\mathrm{L} = \nu_j - \delta_j$), which, in the interaction picture with respect to \cref{eq:nolaserfield} and after a rotating wave approximation, gives~\cite{Leibfried2003,Lee_2005}
\begin{equation}
    \label{eq:H_tuning}
    H^\mathrm{ion}_{\mathrm{t},n} = \Theta_{n,j}' (a_j^\dagger e^{- i \delta_j t} + a_j e^{i \delta_j t})   \ket{n}\bra{n},
\end{equation}
where $\Theta'_{n, j} = \eta g_{n,A,j}^*g_{n,B,j}/\Delta$ is an AC Stark shift with $\eta$ the Lamb-Dicke parameter. $\Theta'_{n, j}$ can be adjusted by varying either the detuning $\Delta$ or the intensity-dependent couplings $g_{n,A,j}$ and $g_{n,B,j}$. We have omitted a Stark shift term from \cref{eq:H_tuning}, which can be made vanishingly small by choosing appropriate laser parameters~\cite{Dietrich2003,Ballance2014}.

Finally, vibronic coupling between states $\ket{n}$ and $\ket{m}$ requires that one of the Raman beams be bichromatic and contain two frequency tones of equal amplitude. The frequency differences between each bichromatic tone and the other Raman beam are $\Delta\omega_\mathrm{L}^\pm = (\omega_{m} - \omega_{n}) \pm (\nu_j + \delta_j) + (\chi_{m} - \chi_{n})$. This interaction (akin to a M{\o}lmer-S{\o}rensen interaction~\cite{Molmer1999}), in the interaction picture with respect to \cref{eq:nolaserfield} and after a rotating wave approximation, gives the coupling Hamiltonian~\cite{Molmer1999,Lee_2005}
\begin{multline}
    H^\mathrm{ion}_{\mathrm{c},{n,m}} = \Omega'_{n, m, j} (a^\dagger_j e^{- i \delta_j t } + a_j e^{i \delta_j t}) \times \\ (\ket{n}\bra{m} e^{- i (\chi_{m} - \chi_n)t}  + \mathrm{h.c.}),
   \label{eq:H_coupling}
\end{multline}
with $\Omega'_{n,m,j} = \eta g_{n,m,A,j}^* g_{n,m,B,j}/2\Delta$, where $g_{n,m,A,j}$ is the light-matter coupling of the monochromatic Raman beam and $g_{n,m,B,j}$ is the light-matter coupling of both tones in the bichromatic Raman beam. $\Omega'_{n,m,j}$ can then be adjusted by varying either $\Delta$, $g_{n,m,A,j}$, or $g_{n,m,B,j}$. 

The electronic-coupling, vibronic-tuning, and vibronic-coupling Hamiltonians can be added for any modes and any electronic states by introducing interactions as described above. For $N$ ions and $d$ electronic states, and after moving to a further interaction picture with respect to $H = \sum_{j=1}^{3N}\delta_j a^\dagger_j a_j + \sum_{n=0}^{d-1} \chi_n \ket{n}\bra{n}$, the overall simulator Hamiltonian becomes
\begin{multline}
    H^\mathrm{sim} = \sum_{j=1}^{3N} \delta_j a^\dg_j a_j 
    + \sum_{n=0}^{d-1} \chi_n \ket{n}\bra{n} 
    + \sum_{ n\neq m} \Omega_{n,m} \ket{n}\bra{m}\\
    + \sum_{n} \sum_{j \in \mathbf{t}} \Theta'_{n,j} (a^\dg_j + a_j) \ket{n}\bra{n} \\
    + \sum_{n \neq m} \sum_{j \in \mathbf{c}} \Omega'_{n,m,j}(a^\dg_j + a_j) \ket{n}\bra{m}.
    \label{eq:closed-system-simulatorH}
\end{multline}

$H^\mathrm{sim}$ is a direct mapping of the molecular Hamiltonian of \cref{eq:closed-system-molecularH}. The simulator is fully programmable: the parameters of \cref{eq:closed-system-simulatorH}---$\delta_j$, $\chi_{n}$, $\Omega_{n,m}$, $\Theta'_{n,j}$, and $\Omega'_{n,m,j}$---can be tuned independently by adjusting suitable laser parameters, such as the intensity or frequency~\cite{MacDonell2021}.

The scaling factor $F$ for molecular simulations on ion traps is typically between $F^{\mathrm{cs}}_{\mathrm{min}} \sim \num{e-12}$ and $F^{\mathrm{cs}}_{\mathrm{max}} \sim \num{e-10}$ (fixed by the maximum achievable laser power for Raman interactions), the ratio between molecular femtosecond timescales and trapped-ion millisecond timescales. The slower ion-trap dynamics, combined with the sub-microsecond timing resolution of trapped-ion simulators, means that greater time resolution is available on an MQB simulator than in direct spectroscopic experiments on a molecule~\cite{Valahu2023}. The long coherence times in trapped-ion systems allow the simulation of chemical dynamics for hundreds of picoseconds~\cite{MacDonell2023}, which is sufficient to observe ultrafast photochemical dynamics.

\begin{figure*}[t]
	\centering
	\includegraphics[width=\textwidth]{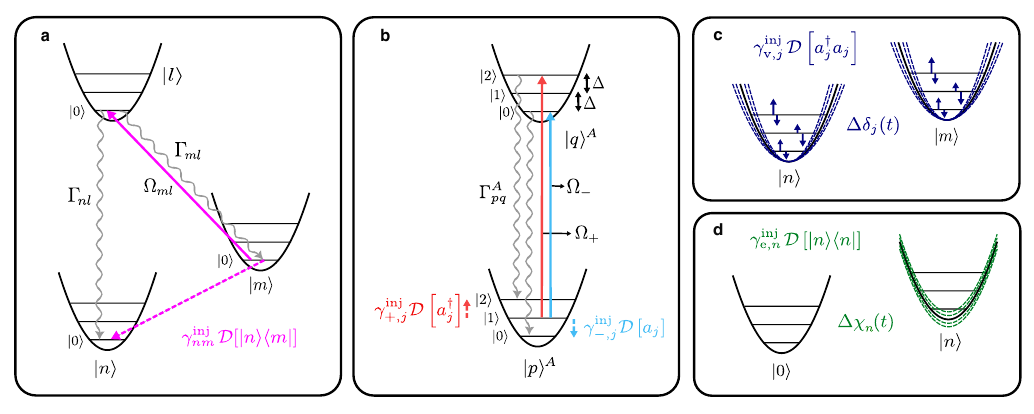}
    \vspace{-6mm}
	\caption{Injecting the four molecular Lindblad dissipators into a trapped-ion MQB simulator. 
    \textbf{a)} Optical pumping to control the radiative electronic relaxation from $\ket{m}$ to $\ket{n}$. The $\ket{m}$ population is pumped (pink solid arrow) to an auxiliary state $\ket{l}$, chosen for its rapid spontaneous decay (wavy arrows) back to $\ket{n}$ and $\ket{m}$. The population at $\ket{m}$ effectively decays to $\ket{n}$ (pink dashed arrow) with rate $\gamma_{nm}^{\mathrm{inj}}$. 
    \textbf{b)} Resolved sideband interactions applied to an ancilla ion ($A$) injecting vibrational heating and cooling. A bichromatic laser with coupling strengths $\Omega_{\pm}$ (red and blue arrows) and detunings $\Delta=\pm \nu_j$ changes the vibrational state from $\ket{i}$ to $\ket{i \pm 1}$. This is followed by an electronic decay from $\ket{q}^A$ to $\ket{p}^A$ at a rate $\Gamma_{pq}^A$ (wavy arrows), without affecting the vibrational state.
    \textbf{c)} Injection of vibrational pure dephasing. A noisy trap electrode or targeted laser field induces fluctuations in the $j$th vibrational mode potential strength (blue dotted potentials), which in turn induce fluctuations $\Delta \delta_j (t)$ in the mode's energy levels (small blue arrows).
    \textbf{d)} Injection of electronic pure dephasing. A noisy magnetic field or targeted laser field induces fluctuations $\Delta \chi_n (t)$ in the electronic energy level $\ket{n}$ (green dotted potentials). 
    } 
    \label{fig:fig2}
\end{figure*}

\section{Open-system MQB simulation on ion traps}

Each chemical dissipator listed in \cref{sec:Lindbladsection} can be engineered on a trapped-ion MQB simulator, using a combination of light-matter interactions and various types of noise injection (summarised in \cref{fig:fig1}). For each case, we derive a rate for the injected dissipation, which we show to be fully controllable, allowing one to tune the simulator dissipator rates of~\cref{eq:inj}.

\subsection{Radiative electronic relaxation}
\label{subsubsec:EleRel}

In an MQB architecture, radiative electronic relaxation corresponds to the native spontaneous decay of an excited state $\ket{m}$ to a lower state $\ket{n}$ at a rate $\Gamma_{nm}$. This is a native dissipative process with Lindblad operator $L_i = \ket{n}\bra{m}$ and rate $ \gamma_{nm}^{\mathrm{nat}} = \Gamma_{nm}$.

When the native $\Gamma_{nm}$ is insufficient to map the radiative electronic relaxation rate of the molecule, optical pumping~\cite{Kastler1950}---a common technique used in operations such as laser cooling~\cite{Cohen1990}, dissipative quantum state preparation~\cite{Barreiro_2011}, and measurement~\cite{Itano1993}---can artificially decrease the lifetime of a given excited state. The procedure is depicted in \cref{fig:fig2}a and involves driving a transition from $\ket{m}$ to an auxiliary high-lying electronic state $\ket{l}$ that can decay to $\ket{n}$ with $\Gamma_{nl} > \Gamma_{nm}$. The driving is achieved using a laser with interaction strength $\Omega_{ml}$ and frequency $\omega_{\mathrm{ext}}$, which is detuned from the electronic transition by $\Delta_{ml} = \omega_{\mathrm{ext}} - (\omega_{l} - \omega_{m})$. The population in $\ket{l}$ decays to $\ket{n}$ and $\ket{m}$ with rates $\Gamma_{nl}$ and $\Gamma_{ml}$, respectively. Population that returns to $\ket{m}$ repeats the optical-pumping cycle until it is completely transferred to $\ket{n}$. The state $\ket{l}$ can be adiabatically eliminated from the dynamics when $\Gamma_{nl} \gg \Omega_{ml}$, resulting in an effective decay rate~\cite{Marzoli_1994}
\begin{equation}\label{eq:sigma_}
        \gamma_{nm}^{\mathrm{inj}} = \frac{\Omega_{ml}^2}{(\Gamma_{nl} + \Gamma_{ml})^2 + 4 \Delta_{ml}^2} \Gamma_{nl}.
\end{equation}
$\gamma_{nm}^{\mathrm{inj}}$ can be tuned over a wide range by varying the laser properties $\Omega_{ml}$ and $\Delta_{ml}$.

Photon scattering during optical pumping imparts momentum kicks to the ions that can heat the vibrational modes. However, this heating rate is negligible in practice, being quadratic in the Lamb-Dicke parameter $\eta$, $\gamma_+ = \Gamma'_{nm} \eta^2$~\cite{Rasmusson2024}. Typical values $\Gamma'_{nm} \approx \SI{0.02}{s^{-1}}$ and $\eta = 0.1$ give $\gamma_+ = \SI{2e-4}{s^{-1}}$, which is negligible compared to typical heating rates in ion-traps, which are in the range $0.1$ to $\SI{10}{s^{-1}}$ (see \cref{fig:fig1}). In any event, what little heating is induced could always be considered as additional useful heating.

The derivation of \cref{eq:sigma_} assumes that population can only decay from $\ket{l}$ to $\ket{n}$ or $\ket{m}$. If there are dipole-allowed transitions to other electronic states $\ket{m'}$, additional driving beams can repump the population from $\ket{m'}$ back into the optical pumping cycle~\cite{Cornelius_thesis}.

\begin{figure*}[t]
	\centering
	\includegraphics[width=\textwidth]{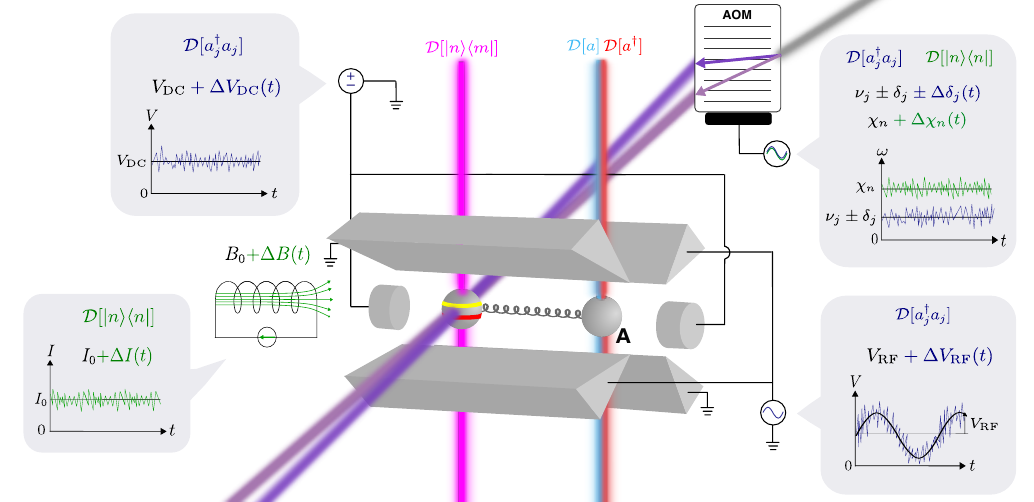}
	\caption{Simulating MQB open-system molecular dynamics in an ion trap. The ions (spheres) are trapped in vacuum in a linear chain by two DC (end caps) and four RF electrodes (triangular blades). The closed-system simulation is achieved using a qudit ion representing the molecular electronic states (yellow $\ket{m}$ and red $\ket{n}$ lines) and collective ion vibrational modes (grey spring) representing molecular vibrations. Vibronic coupling is simulated with a bichromatic laser field (two shades of purple) originating from a single incident laser (grey) modulated by an AOM. An open-system simulation requires an ancilla ion (A)  (which shares vibrational modes with the qudit ion) and controllable implementations of the four Lindblad dissipators: $\mathcal{D}[\ket{n}\bra{m}]$ (pink) can be injected via optical pumping; $\mathcal{D}[a_j]$ (light blue) and $\mathcal{D}[a^\dg_j]$ (red) can be achieved using blue- and red-sideband interactions; $\mathcal{D}[\ket{n}\bra{n}]$ (green), can be injected in two ways: (1) using a noisy detuning ($\Delta \chi_n$) in the vibronic-coupling laser or (2) using a noisy current $\Delta I$ to fluctuate the magnetic field ($B_0 +\Delta B$); and $\mathcal{D}[a^\dg_j a_j]$ (dark blue) can also be implemented in two ways: (1) using a fluctuating detuning $\Delta \nu_j$ in the vibronic-coupling laser or (2) using noisy voltages $\Delta V_{\mathrm{RF}}$ and $\Delta V_{\mathrm{DC}}$ on the electrodes.
    }
	\label{fig:fig3}
\end{figure*}

\subsection{Vibrational heating and cooling}
\label{subsec:VibRel}

Vibrational heating and cooling in a trapped-ion simulator commonly occur due to electric-field noise. This dissipation is routinely characterised experimentally~\cite{Brownnutt2015,Bruzewicz2015} and is described by Lindblad dissipators $\mathcal{D}[a]$ and $\mathcal{D}[a^\dag]$. The corresponding rates are approximately, $\gamma_{+,j}^{\mathrm{nat}} = \gamma_{-,j}^{\mathrm{nat}} \exp{(-\nu_j / k_B T^{\mathrm{nat}})} \approx \gamma_{-,j}^{\mathrm{nat}}$, because typical trap frequencies ($\nu_j/2\pi \sim \SI{1}{MHz}$) are much smaller than $k_B T^{\mathrm{nat}}$ for $T^{\mathrm{nat}}$ between 4 and \SI{300}{K}. These native rates vary from 0.1 to 1000~quanta/s~\cite{Talukdar2016,Bruzewicz2015,Turchette2000}. If the native dissipative rates are insufficient, additional dissipation can be injected.

\subsubsection*{Single-mode heating and cooling}

The injection of vibrational heating and cooling was proposed for ion-trap simulators in the original MQB paper~\cite{MacDonell2021}. The proposed injection was via resolved-sideband laser interactions, where the desired molecule-environment coupling and temperature could be controlled by tuning the laser parameters~\cite{MacDonell2021}. However, these laser interactions could change the ion's electronic state, disturbing the ongoing coherent simulation. 

To overcome this limitation, a similar dissipation scheme can be implemented using an ancillary ion that shares the motion with the MQB qudit ion (i.e., sympathetic cooling~\cite{Leibfried2003,Larson1986,so2024}), without affecting the electronic states of the qudit. Dissipators $\mathcal{D}[a_j]$ and $\mathcal{D}[a^\dagger_j]$ can then be engineered using a bichromatic laser (one with two frequency tones), interacting with the ancilla ion. The frequency of each tone is set so that it sympathetically cools or heats a shared vibrational mode. The ratio of the two tones' strengths can be adjusted to simulate a desired environment's temperature.

\Cref{fig:fig2}b depicts the electronic transition $\ket{p}^A \to \ket{q}^A$ in the ancillary ion $A$ driven by a bichromatic laser. The interaction strengths are $\Omega_-$ and $\Omega_{+}$, with detunings $\Delta = -\nu_j$ for the red sideband and $\Delta = +\nu_j$ for the blue sideband, respectively. The electronic population decays from $\ket{q}^A$ to $\ket{p}^A$ at rate $\Gamma_{pq}^A$ (this can be due to spontaneous emission or engineered as in \cref{eq:sigma_}). The dynamics of the vibrational mode is then described by the Lindblad master equation~\cite{Stenholm_1986, Cirac_1992}
\begin{equation}\label{eq:ionLindblad}
    \frac{d\rho_j}{dt} = -i [ \nu_j a^\dg_j a_j, \rho_j ] + \left( \gamma_{+,j}^{\mathrm{inj}} \mathcal{D} [ a^\dg_j ] + \gamma_{-,j}^{\mathrm{inj}} \mathcal{D} [a_j] \right)\rho_j,
\end{equation}
where $\rho_j$ is the reduced density matrix of mode $j$. In the limit $\Omega_{\pm} \ll \Gamma_{pq}^A$ (which usually holds in trapped-ion simulators), the injected dissipation rates depend only on the laser parameters~\cite{Stenholm_1986,Cirac_1992,Marzoli_1994}:
\begin{align}\label{eq:Apm}
    \gamma_{\pm,j}^{\mathrm{inj}} &= \eta^2_j \Gamma_{pq}^A \left[ B(\Delta \mp \nu_j) + \alpha B(\Delta) \right],\\
    B(\Delta) &= \frac{\Omega_{\pm}^2}{\left(\Gamma_{pq}^A\right)^2 + 4\Delta^2 },
\end{align}
where $\eta_j$ is the Lamb-Dicke parameter and $\alpha$ is an angular factor ($\alpha = 2/5$ for dipole transitions~\cite{Marzoli_1994}). 

In molecules, heating and cooling most commonly arise when the molecule is coupled to a thermal environment. Simulating a thermal environment requires that the heating and cooling rates satisfy detailed balance, $\gamma_{+,j}^{\mathrm{sim}} = \gamma_{-,j}^{\mathrm{sim}} \exp(- \nu_j/k_\mathrm{B}T^{\mathrm{sim}})$, where $T^{\mathrm{sim}} = (\nu_j/\nu_j^{\mathrm{mol}}) T^{\mathrm{mol}}$. In terms of the injected rates,
\begin{equation}
    \gamma_{+,j}^{\mathrm{inj}} = \gamma_{-,j}^{\mathrm{inj}}\zeta^2 - \gamma_{-,j}^{\mathrm{nat}}(1-\zeta^2),
\end{equation}
where $\zeta^2 = \exp(- \nu_j^{\mathrm{mol}}/k_\mathrm{B}T^{\mathrm{mol}})$. If, as is the case in good ion traps, $\gamma_{-,j}^{\mathrm{nat}} \ll \gamma_{-,j}^{\mathrm{inj}}$, the Boltzmann factor can be incorporated into the simulation through the interaction strength of \cref{eq:Apm} by setting $\Omega_{+} = \zeta \Omega_{-}$. In other words, the environment temperature $T^{\mathrm{mol}}$ can be set by tuning the ratio between the blue- and red-sideband interaction strengths, while their absolute value determines the strength of the system-environment coupling.

\subsubsection*{Global vibrational heating and cooling}
\label{sec:GlobalHeatCool}

Using mode-resolved laser interactions as described above means that the number of required laser interactions scales linearly with the number of modes, a cost that can be significantly reduced if all the heating and cooling rates are similar. In many cases, the precise spectral density of the environment is either unknown or unimportant; in those cases, it is common to assume that the heating rates for all vibrational modes are equal (and likewise for the cooling rates). In those cases, where a precise spectral density is not required, similar heating and cooling rates can be injected for all modes using a single broadband laser that can heat and cool all modes simultaneously. According to \cref{eq:Apm}, this can be achieved if the laser parameters ($\Gamma_{pq}^{A}$, $\Omega_{\pm}$, $\Delta$) and the vibrational frequencies ($\nu_j$) are similar for all vibrational modes. A global laser interaction with a sufficiently broadband spectrum can target all modes, setting the same values for $\Gamma_{pq}^{A}$ and $\Omega_{\pm}$, as well as small $\Delta$ for all modes.

\subsection{Electronic and vibrational pure dephasing}
\label{sec:Puredephasingcontrolled}

Electronic and vibrational dephasing can be injected by inducing classical stochastic fluctuations in the energy levels of the electronic and vibrational states. The resulting dephasing rates can be tuned by controlling the variances and correlation times of the fluctuations.

We consider a fluctuating Hamiltonian proportional to an arbitrary operator $O$,
\begin{equation}
    \label{eq:Hamiltonian_arb_op_noisy}
    H(t) = \Delta\Phi(t) \, O,
\end{equation}
where the fluctuations, $\Delta\Phi(t)$, are a zero-mean stochastic process. To simulate pure dephasing, we consider the ensemble average of the density matrix $\langle \rho \rangle$ over many realisations of $\Delta\Phi(t)$~\cite{Chenu2017,Gu2019}. This approach reflects experimental quantum simulations, where expectation values are measured by averaging over many outcomes. The resulting dissipator is~\cite{multilevelPD, multilevelPDOG}
\begin{align}
    \mathcal{D}[O]\langle\rho\rangle = \gamma \left(O \langle \rho \rangle O^\dagger - \tfrac{1}{2} \{O^\dagger O, \langle \rho \rangle\}\right), 
\end{align}
with rate $\gamma = \frac{1}{2} \int^t_0 \langle \Delta\Phi(t) \Delta\Phi(s) \rangle \, ds$, where $\langle \Delta\Phi(t)\Delta\Phi(s)\rangle$ is the auto-correlation function of the fluctuations. For Gaussian, Markovian, and wide-sense stationary noise, the rate simplifies to~\cite{Yang2016}
\begin{equation}\label{eq:dephasing_rate}
    \gamma = \tfrac{1}{2} \langle \Delta\Phi(t)^2\rangle \tau_c,
\end{equation}
where $\langle \Delta\Phi(t)^2\rangle$ is the variance of the noise and $\tau_c \ll 1/\gamma$ is its correlation time. Either of these quantities can be used to adjust the dephasing rate. 

In the following, we present several ways to inject stochastic fluctuations into the terms of the simulator Hamiltonian of \cref{eq:closed-system-simulatorH}, in a way that results in controllable electronic and vibrational dephasing.

\subsubsection*{Dephasing single modes or electronic levels by noise}

Vibrational and electronic pure dephasing can be independently engineered by adding frequency fluctuations to the laser beams that drive the closed-system dynamics. Such frequency fluctuations can be imprinted on the beams by frequency modulating the acousto-optical modulator (AOM) already used for the closed-sysem simulation. (\cref{fig:fig3}). Thus, no additional experimental hardware is necessary to inject pure dephasing.

Vibrational pure dephasing of mode $j$ is implemented by injecting frequency fluctuations into both vibronic Hamiltonians of \cref{eq:H_tuning,eq:H_coupling}, with the replacement $\delta_j \rightarrow \delta_j + \Delta\delta_j(t)$. Electronic pure dephasing of qudit levels $n$ and $m$ requires adding frequency fluctuations to the coupling Hamiltonians of \cref{eq:H_coupling,eq:H_coupling_electronic}, with the replacements $\chi_n \rightarrow \chi_n + \Delta \chi_n(t)$ and $\chi_m \rightarrow \chi_m + \Delta \chi_m(t)$. To this end, we modify the frequency differences of the Raman beams as follows. For the vibronic tuning interaction, we set $\Delta\omega_\mathrm{L} = \nu_j - (\delta_j + \Delta\delta_j(t))$. For the vibronic coupling interaction, we set $\Delta\omega_\mathrm{L}^\pm = (\omega_m - \omega_n) \pm (\nu + \delta_j + \Delta\delta_j(t)) + (\chi_m + \Delta\chi_{m}(t)) - (\chi_n - \Delta\chi_n(t))$. For the electronic coupling interaction, we set $\Delta\omega_\mathrm{L} = (\omega_m  - \omega_n) + (\chi_m + \Delta\chi_m(t)) - (\chi_n + \Delta \chi_n(t))$. Moving to an interaction picture with these detunings, the simulation Hamiltonian becomes
\begin{multline}
    H^\mathrm{sim} =  \sum_{j=1}^{3N} (\delta_j + \Delta\delta_j(t)) a^\dg_j a_j + \\
    \sum_{n=0}^{d-1} (\chi_n + \Delta\chi_n(t)) \ket{n}\bra{n} + \sum_{n\neq m} \Omega'_{n, m} \ket{n}\bra{m} + \\
    \sum_{n} \sum_{j \in \mathbf{t}} \Theta'_{n,j} (a^\dg_j + a_j) \ket{n}\bra{n} + \\
    \sum_{n \neq m} \sum_{j \in \mathbf{c}} \Omega'_{n,m,j}(a^\dg_j + a_j) \ket{n}\bra{m}.
\end{multline}

$H^\mathrm{sim}$ contains stochastic terms of the form in~\cref{eq:Hamiltonian_arb_op_noisy}, from which we can retrieve the induced dephasing rates. Vibrational dephasing for mode $j$ is obtained by setting $O = a_j^\dagger a_j$ and $\Delta\Phi(t) = \Delta\delta(t)$, resulting in rate $\gamma_{\mathrm{v},j}^\mathrm{inj} = \langle \Delta\delta_j(t)^2\rangle \tau_c/2$. Similarly, electronic dephasing involving qudit level $n$ is obtained by setting $O = \ket{n}\bra{n}$ and $\Delta\Phi(t) = \Delta\chi_n(t)$, resulting in rate $\gamma_{\mathrm{e},n}^\mathrm{inj} = \langle \Delta\chi_n(t)^2\rangle \tau_c/2$. The total electronic dephasing rate between levels $n$ and $m$ is then $\gamma_{nm}^\mathrm{inj} = \gamma_{\mathrm{e},n}^\mathrm{inj} + \gamma_{\mathrm{e},m}^\mathrm{inj}$. Because the magnitudes of the fluctuations and their correlation times are independently controlled, the dephasing rate of each mode and electronic level is independently programmable.

When engineering dephasing, large frequency fluctuations in the laser fields may cause unwanted off-resonant couplings to other vibrational modes. These couplings can be minimized by increasing $\tau_c$, which leads to a smaller variance of the noise, or by choosing ion trap parameters which increase the frequency spacings between the vibrational modes.

\subsubsection*{Global electronic dephasing by magnetic field fluctuations}
\label{sec:global_dephasing}
   
A simplified injection of electronic pure dephasing can be implemented if all electronic states involved in the simulation are affected by global stochastic frequency fluctuations, creating similar pure dephasing rates. As in \cref{sec:GlobalHeatCool}, this scheme can be used to simulate open systems for which there is not enough information about the pure dephasing rate of each electronic state. Assuming that different electronic states experience similar pure dephasing rates can be a satisfactory approximation in these cases.  

Global dephasing can be implemented using fluctuating magnetic fields $\Delta B$, which alter the energies of magnetically sensitive electronic states. To first order in $\Delta B$, the fluctuation in the $n$th electronic energy is~\cite{whineland1998}
\begin{equation}
    \Delta\omega_n = \left( \diff{\omega_n}{B} \right)_{\!B_0} \!\! \Delta B,
    \label{eq:Bfluctuations}
\end{equation}
where $(d\omega_n/dB)_{B_0}$ is the sensitivity at the average field $B_0$. The fluctuation $\Delta B$ can be engineered using a noisy current $\Delta I$ in the magnetic field source, such that $\Delta B = (dB/dI) \Delta I$. For cylindrical solenoids, the typical source of magnetic fields in ion traps, $dB/dI = C\mu_0/L$, where $C$ is the number of windings, $L$ is the length of the solenoid, and $\mu_0$ is the vacuum permeability. Using \cref{eq:Bfluctuations}, \cref{eq:dephasing_rate} gives the injected dephasing rate,
\begin{equation}
    \gamma_{\mathrm{e},n}^{\mathrm{inj}} = \frac{1}{2} \left(\frac{d\omega_n}{dB} \frac{dB}{dI} \right)^2 \langle \Delta I^2 \rangle \tau_c. 
\end{equation}
For these rates to be similar, one needs to find electronic states with similar $d\omega_n/dB$. For states that are linearly sensitive to the magnetic field, $d\omega_n/dB$ is proportional to the projection of the total angular momentum onto the quantisation axis ($m_J$), which allows states of similar $m_J$ to be selected as the qudit states. In particular, hyperfine states within the same angular momentum manifold experience similar sensitivities~\cite{Breit1931}.

\subsubsection*{Global vibrational dephasing by trapping voltage fluctuations}
\label{sec:glo_vib_dep}

Similarly to electronic dephasing in the previous section, the same (or similar) vibrational pure dephasing rates are often assumed to affect all modes in an open molecule~\cite{Schneider_1990, Stock_1990, Schneider_1989}. This global dephasing can be implemented by injecting noise into the strength of the ions' confining potentials, which control the vibrational mode frequencies. This injection can be achieved by inducing voltage fluctuations in the electrodes that create the confining electric fields (\cref{fig:fig3}).

The ions' vibrational frequencies are determined by the three-dimensional electric fields produced by the four radio-frequency (RF) and two static (DC) electrodes (\cref{fig:fig3}). The blade electrodes, with RF signal of peak voltage $V_\mathrm{RF}^{(0)}$ produce a radially confining potential (in the $xy$ plane), while the end caps with DC voltage $V^{0}_\mathrm{DC}$ confine the ions axially (along $z$). A chain of $N$ ions has $3N$ vibrational modes, of which the radial frequencies $\nu_{r,i}$ are determined by the RF voltage, while the axial frequencies $\nu_{z,i}$ by the DC voltage. One can relate the vibrational frequencies to the center-of-mass frequencies $\nu_{r,0}$ and $\nu_{z,0}$ using geometric factors: $\nu_{r,i} = \kappa_{r,i} \nu_{r,0}$ and $\nu_{z,i} = \kappa_{z,i} \nu_{z,0}$~\cite{whineland1998}, where $\kappa_{r,i}$ and $\kappa_{z,i}$ are typically of the same order of magnitude for all $i$~\cite{James1998}.

Small voltage fluctuations change the vibrational frequencies according to the corresponding sensitivities:
\begin{align}
    \Delta\nu_{r,i} &= \left( \diff{\nu_{r,i}}{V_\mathrm{RF}} \right)_{V_\mathrm{RF}^{(0)}} \!\! \left(V_\mathrm{RF}-V_\mathrm{RF}^{(0)}\right), \\
    \Delta\nu_{z,i} &= \left( \diff{\nu_{z,i}}{V_\mathrm{DC}} \right)_{V^{(0)}_\mathrm{DC}} \!\! \left(V_\mathrm{DC}-V^{(0)}_\mathrm{DC}\right).
    \label{eq:Vfluctuations}
\end{align}
We therefore obtain the injected pure dephasing rates,  
\begin{align}
    \gamma_{\mathrm{v},r,i}^{\mathrm{inj}} & = \frac12 \kappa_{r,i}^2 \left(\frac{d\nu_{r,}}{dV_\mathrm{RF}}\right)^2 \langle\Delta V_{\mathrm{RF}}^2\rangle \tau_c, \\
    \gamma_{\mathrm{v},z,i}^{\mathrm{inj}} & = \frac 12 \kappa_{z,i}^2 \left(\frac{d\nu_{z,0}}{dV_\mathrm{DC}}\right)^2 \langle\Delta V_{\mathrm{DC}}^2\rangle \tau_c.
\end{align}
Therefore, global pure vibrational dephasing can be implemented if all of the geometric factors $\kappa$ are sufficiently close to each other. The variances of the pure dephasing of axial and radial modes can be separately tuned by varying the amplitudes of the DC and RF voltage fluctuations, respectively.

\renewcommand{\tabcolsep}{4pt}
\renewcommand{\arraystretch}{1.2}
\sisetup{table-number-alignment = left, table-align-text-after = false}
\begin{table*}[t]
	\centering
     \begin{tabular}{ l S[table-format=2.2] S[table-format=3.2] S[table-format=4.1] l S[table-format=2.4] S[table-format=3.2] l}
         \toprule
            & & \multicolumn{3}{c}{Triiodide $\left(F=\num{1.6e-11}\right)$} & \multicolumn{3}{c}{Pyrazine $\left(F=\num{2.6e-11}\right)$} \\
        \cmidrule(r){3-5} \cmidrule{6-8}
           Dissipation & {$\gamma_i^{\mathrm{nat}}/\unit{s^{-1}}$} & {$\gamma_i^{\mathrm{mol}}/\unit{ps^{-1}}$} & {$\gamma_i^{\mathrm{sim}}/\unit{s^{-1}}$} & Implementation & {$\gamma_i^{\mathrm{mol}}/\unit{ps^{-1}}$}  & {$\gamma_i^{\mathrm{sim}}/\unit{s^{-1}}$} & Implementation \\
         \midrule
           Elec.\ relaxation & 0 & 0 & 0 & $\Omega_{ml}/2\pi = 0$ & 0 & 0 & $\Omega_{ml}/2\pi = 0$ \\
           Vib.\ cooling & 0.20 {~\cite{Valahu2023}} & 0.52 {~\cite{Banin_1994}} & 8.6 & $\eta\Omega_-/2\pi = \SI{11}{kHz}$ & 1.0 {~\cite{Schneider_1990}} & 27 & $\eta\Omega_-/2\pi = \SI{20}{kHz}$ \\
           Vib.\ heating & 0.20 {~\cite{Valahu2023}} & 0.31 {~\cite{Banin_1994}} & 5.0 & $\eta\Omega_+/2\pi = \SI{8.5}{kHz}$ & 0.0077  {~\cite{Schneider_1990}} & 0.20 & $\eta\Omega_+ = 0$ \\
           Vib.\ dephasing & 29 {~\cite{Valahu2023}} & 1.8 {~\cite{Banin_1994}} & 29 & $\sqrt{\braket{\Delta \delta_j^2}}/2\pi= 0$ & 2.1  {~\cite{Stock_1990}} & 55 & $\sqrt{\braket{\Delta \delta_j^2}}/2\pi = \SI{2.3}{kHz}$ \\
           Elec.\ dephasing & 0.12 {~\cite{Tan2023}} & 120 {~\cite{Banin_1994}} & 2000 & $\sqrt{\braket{\Delta \chi_n^2}}/2\pi  = \SI{20}{kHz}$ & 33 {~\cite{Schneider_1989}} & 870 & $\sqrt{\braket{\Delta \chi_n^2}}/2\pi = \SI{13}{kHz}$ \\
 	    \bottomrule
     \end{tabular}
	\caption{Dissipation rates $\gamma_i^{\mathrm{mol}}$ and the corresponding simulated rates $\gamma_i^{\mathrm{sim}} = F \gamma_i^{\mathrm{mol}}$ in an ion-trap MQB simulator for triiodide in solution~\cite{Banin_1994} and for a 3-mode, 2-state pyrazine molecule where the dissipation is due to the other vibrational modes. The column ``Implementation'' denotes experimental parameters for implementing $\gamma_i^{\mathrm{sim}}$.
    }
	\label{tab:I3}
\end{table*}

\section{Examples}
\label{sec:Examples}

We apply our method to two well-studied examples of open-molecular systems: (1) triiodide in a polar solvent, and (2) pyrazine modelled as an LVC system with dissipative dynamics due to its intramolecular vibrational modes. In both examples, we use molecular Lindblad dissipation rates determined in earlier works and map them to equivalent rates in an experimentally realistic trapped-ion MQB simulator.

Our approach requires predetermined Lindblad dissipation rates for the open system of interest, which are obtainable from time-resolved spectroscopy, simulations on classical computers, or a combination of both. For example, time-resolved spectroscopic experiments often report population decay and dephasing times $T_1$ and $T_2$, respectively, for both electronic states and vibrational modes. The electronic pure dephasing rate is then given by ${\gamma_{\mathrm{e},n}^{\mathrm{mol}}} = 1/T_{2\mathrm{e}} - 1/2T_{1\mathrm{e}}$, where the electronic dephasing time $T_{2\mathrm{e}}$ can be obtained from fitting the electronic time-dependent spectra~\cite{Stock_1990}. $T_{1\mathrm{e}}$ is extracted from the radiative quantum yield $Y_R = T_{1\mathrm{e}} \gamma_{nm}^{\mathrm{mol}}$~\cite{Stock_1990}, where the radiative electronic relaxation rate is determined by the strength of the transition dipole moment $\ket{m} \to \ket{n}$~\cite{Stock_1990}. The analogous vibrational times $T_{2\mathrm{v}}$ and $T_{1\mathrm{v}}$---also obtainable from time-dependent spectroscopy---determine $\gamma_{-,j}^{\mathrm{mol}} - \gamma_{+,j}^{\mathrm{mol}} = 1/T_{1\mathrm{v}}$ and $\gamma_{\mathrm{v},j}^{\mathrm{mol}} = 1/T_{2\mathrm{v}} - 1/2T_{1\mathrm{v}}$~\cite{Banin_1994}.

\subsection*{Triiodide in a polar solvent} 

Triiodide gained interest after time-resolved spectroscopic experiments of its photodissociation dynamics in the condensed phase showed the vibrational decoherence effects of solvents~\cite{Banin_1994, Banin_1992, Banin1993}. In particular, the key role of the symmetric stretch mode $\nu_1$ at \SI{112}{cm^{-1}} makes it ideal for probing solvent effects on the dynamics. 

The relevant dissipation rates were measured by pumping the molecule to a dissociative electronic surface and then monitoring the solvent-induced vibrational dissipation using time-delayed probe pulses. The resulting spectrum was fitted to a decaying sine function to obtain the vibrational relaxation and dephasing times $T_{1\mathrm{v}}$ and $T_{2\mathrm{v}}$~\cite{Banin_1994, Gershgoren2003}. The electronic dephasing rate was also chosen to fit the spectral modulations. The ranges of the corresponding dissipation rates obtained for water and ethanol solvents are given in \cref{fig:fig1}. Radiative electronic relaxation was not included because it is much slower than the other processes.

These measured rates can be mapped onto an MQB simulator. \Cref{tab:I3} shows the average of the dissipative rates in water and ethanol and gives examples of typical native dissipation rates $\gamma_i^{\mathrm{nat}}$ for an \ce{Yb+} trapped-ion simulator~\cite{Valahu2023, MacDonell2023}. We select the scaling factor $F$ as described in \cref{sec:SimulationFramework} to minimise the injected dissipation. In this example, vibrational pure dephasing determines $F = \gamma_{\mathrm{v},r}^{\mathrm{nat}}/\gamma_{\mathrm{v},1}^{\mathrm{mol}} =\num{1.6e-11}$, so that no injection of this dissipation type would be required. $F$ is then used to scale the other rates $\gamma_i^{\mathrm{sim}}$ in \cref{tab:I3}. Electronic and vibrational dephasing, occurring on timescales of tens and hundreds of femtoseconds, respectively, would affect the MQB dynamics on 1--30~ms timescales. Vibrational dissipation would be mapped from a picosecond in the molecule to $\SI{0.1}{s}$ in the ion trap.

All calculated parameters for simulating $\gamma_{i}^{\mathrm{sim}}$ are within existing experimental capabilities, for which we have used representative values from past ion-trap MQB simulations~\cite{MacDonell2021, MacDonell2023, Valahu2023}. The interaction strength $\eta \Omega_{-}/2\pi = \SI{11}{kHz}$, and therefore also the smaller $\Omega_+$, are well within the range achievable by tuning the laser power~\cite{Baldwin2021}. $\Omega_{-}$ is calculated from \cref{eq:Apm}, using $\Gamma_{pq}^A/2\pi = \SI{20}{MHz}$ and $\nu_x/2\pi = \SI{1.34}{MHz}$, values reported for an Yb$^+$ ion trap~\cite{Valahu2023}. $\Omega_{+}$ is obtained from the Boltzmann factor using $\nu_1 = \SI{112}{cm^{-1}}$ and $T^{\mathrm{mol}} = \SI{300}{K}$~\cite{Banin_1994}. The fluctuations for injecting electronic dephasing, $\sqrt{\braket{\Delta \chi_n^2}}/2\pi = \SI{20}{kHz}$, could be imprinted on the frequency of each Raman beam using an AOM. The electronic dephasing rate $\gamma_{\mathrm{e},n}^{\mathrm{sim}}$ is determined using \cref{eq:dephasing_rate}, where the correlation time $\tau_c$ must be chosen much smaller than the dephasing timescale; we choose $\tau_c = \SI{10}{\micro s} \ll  1/\gamma_{\mathrm{e},n}^{\mathrm{sim}} = \SI{500}{\micro s}$. This choice is well within the capabilities of modern control electronics, which achieves temporal resolutions of \SI{100}{ns}.

\subsection*{Pyrazine} 

Pyrazine is widely used for studying ultrafast nonadiabatic dynamics due to the conical intersection between its excited electronic states~\cite{raab1999a}. This 24-mode molecule has been successfully modelled as an open quantum system in which three vibrational modes, one coupling ($\nu_{10a}$) and two tuning ($\nu_{1}$ and $\nu_{6a}$), are coupled to the first two excited electronic states, $n\pi^*$ and $\pi\pi^*$, while the remaining 21 modes form a weakly coupled environment that leads to dissipative dynamics~\cite{Schneider_1989}. 

As in the triiodide example, dephasing and relaxation constants for electronic states and vibrational modes were retrieved from fitting the experimental absorption spectrum~\cite{Stock_1990,Schneider_1989,Domcke2002,RatnerLindblad2005}. Radiative electronic relaxation from $\pi\pi^*$ to $n\pi^*$ is forbidden and not included. The pure dephasing constant for the state $\pi\pi^*$ was $T_{2,\pi\pi^*}^* = \SI{30}{fs}$. We assume that the constants $T_{1\mathrm{v}} = \SI{1}{ps}$~\cite{Domcke2002,RatnerLindblad2005} and $T_{2\mathrm{v}} = \SI{320}{fs}$~\cite{Schneider_1989} are the same for all modes. 

Using the procedure above, we converted the dissipation times into the Lindblad rates and mapped them to the simulator dissipative rates (\cref{tab:I3}). In particular, we use the tuning mode $\nu_{1} = \SI{1016}{cm^{-1}}$ to exemplify the Lindblad rates for the vibrations. In this case, $F$ comes from the vibrational heating of this tuning mode, $F = \gamma_{+,r}^{\mathrm{nat}}/\gamma_{+,1}^{\mathrm{mol}} =\num{2.6e-11}$, which, along with the $\nu_{6a}$ and $\nu_{10a}$ modes, would not require dissipation injection, i.e., $\Omega_+ = 0$. The other Lindblad rates are scaled as before, $\gamma_i^{\mathrm{sim}} = F\gamma_i^{\mathrm{mol}}$. The experimental parameters for dissipation injection (including single-mode schemes) are calculated as in the triiodide example and reported in \cref{tab:I3}. In this case, vibrational cooling could also be implemented globally by targeting the vibrational modes with one broadband laser interaction, such that the laser power is adjusted to $\Omega_- = \SI{20}{kHz}$. The global vibrational dephasing scheme would also be suitable, with fluctuations $\sqrt{\braket{\Delta \nu_r^2}}/2\pi = \SI{2.3}{kHz}$ produced via trapping potential fluctuations $\Delta V_{\mathrm{RF}}$.

\section{Discussion}

Our approach extends molecular MQB simulation from closed systems to open ones, the latter characterised by the most common set of Lindblad dissipators used in chemistry. Dissipation can be included in the simulation using a combination of native and injected dissipation, an approach that turns native dissipation into a valuable resource for analog simulation. In addition, injected dissipation, described for trapped-ion systems, could be readily implemented using available simulators with few additional hardware modifications and independently of the molecular size.

Using native noise as a simulation resource diminishes the common argument that the accumulation of errors makes analog simulators impractical. Dissipation and decoherence are the main sources of error in analog devices---quantum or classical---and limit the simulation time. However, analog classical simulations were, despite errors, essential for classical computation before large-scale digital computers~\cite{Karplus}. Similarly, successful quantum simulations of closed systems~\cite{Valahu2023,MacDonell2023} have shown that available decoherence times in analog quantum simulators are sufficiently low to allow accurate simulations, up to hundreds of femtoseconds on molecular timescales, which are the most relevant for photochemistry. Here, we have shown that harnessing a simulator's native noise as a resource \textit{always} extends the possible simulation times, up to picoseconds on molecular timescales. 

The result of using noise as a resource is counterintuitive: open-system simulation, which is usually a \textit{harder} problem than closed-system simulation, becomes an \textit{easier} problem on an MQB simulator. On both classical and quantum digital computers, simulating open-system dynamics requires more resources, either to represent density matrices instead of wavefunctions or to represent the environment (or both). By contrast, on an analog simulator, if some of the noise is used for the simulation, the remaining unusable dissipation has a smaller deleterious effect, extending the possible simulation duration. 

The transition from a closed to an open simulation requires minimal additional experimental cost on a trapped-ion MQB simulator. When using Raman lasers to inject dissipation, no additional hardware is required at all, since those lasers are required for the LVC simulation as well. For the other hardware we describe, such as the magnetic-field solenoid, the overhead is constant regardless of the system size (e.g., only one solenoid is needed for any molecule).

Ion traps are particularly versatile MQB simulators. The two example systems show that our approach can accommodate a wide range of molecular systems: from a small inorganic ion to a neutral organic molecule; from a polar solvent to an environment of intramolecular vibrations; different numbers of vibrational modes, with frequencies varying by an order of magnitude; global- and single-mode dissipation injection; and scaling-factor selection to minimise noise injection. 

Like any LVC simulation, our scheme requires a parametrisation of the Hamiltonian in \cref{eq:closed-system-molecularH}. As is standard in non-adiabatic dynamics, the parameters can be obtained using a prior electronic-structure calculation. In addition, we require values of the dissipation rates; these can also be obtained from a prior calculation or by comparison with experimental results on related molecules. For systems without enough experimental information and unknown dissipative rates, our global injection schemes offer an initial approach for estimation of the molecular dissipation effects and for the interpretation of experimental data. For example, our algorithm could be used in the quantum time-domain spectroscopy algorithm~\cite{MacDonell2023} to predict or explain spectral peak broadening in open molecular systems. 

We expect that our approach can be extended to dissipation models more general than the Lindblad master equation. For example, non-Markovian environments include strong coupling between the molecule and the environment with long-time correlation functions~\cite{Breuer2016}. One possibility for engineering a non-Markovian environment could be to inject correlated (coloured) classical noise~\cite{Costa2017} into the amplitude and phase of the laser beams used for vibronic and electronic coupling. Alternatively, a hybrid approach could be pursued, which combines our Lindblad techniques with an explicitly implemented, strongly coupled environment, implemented using additional ancillary ions~\cite{sun2024,so2024} or vibrational modes (with structured spectrum).

\section{Conclusion}

Overall, including native dissipation relaxes the experimental complexity when transitioning from closed- to open-system simulation and reduces error accumulation, allowing for longer, more accurate analog simulations of molecular dynamics. When injected dissipation is necessary, minimal hardware modifications are required. Because of the difficulty of simulating open molecular systems both classically and on digital quantum computers, these advantages position analog quantum simulation as a contender for achieving quantum advantage on a problem of practical importance.

\begin{acknowledgments}
We were supported by the Australian Research Council (FT220100359, FT230100653), the U.S. Office of Naval Research Global (N62909-24-1-2083), the U.S. Army Research Office Laboratory for Physical Sciences (W911NF-21-1-0003), the U.S. Air Force Office of Scientific Research (FA2386-23-1-4062), the Wellcome Leap Quantum for Bio program, the Sydney Quantum Academy (VCOA, MJM), the University of Sydney Postgraduate Award scholarship (VGM), the Australian Government Research Training Program (FS), the National Science and Engineering Research Council of Canada (RJM), the Sydney Horizon Fellowship, Lockheed Martin, and H.\ and A.\ Harley.
\end{acknowledgments}

\bibliography{bib}

\end{document}